\documentclass[preprint]{acm_proc_article-sp}
\usepackage{epsfig}
\newcommand{\UPLB}{University of the Philippines Los Ba\~{n}os}

\begin{document}

\title{A Framework for\\a Multiagent-based Scheduling of\\Parallel Jobs}
\numberofauthors{1}
\author{
\alignauthor Jaderick P. Pabico\\
   \affaddr{Institute of Computer Science}\\
   \affaddr{College of Arts and Sciences}\\
   \affaddr{\UPLB}\\
   \affaddr{College 4031, Laguna, Philippines}\\
   \affaddr{63-49-536-2313}\\
   \email{jppabico@uplb.edu.ph}
}
\date{}
\toappearbox{Proceedings (CDROM) of the 6th Philippine Computing Science Congress (PCSC 2006), Ateneo de Manila University, 28--29 March 2006. pp. C7. \copyright 2006 Computing Society of the Philippines.}
 
\maketitle

\begin{abstract}
This paper presents a multiagent approach as a paradigm for scheduling parallel jobs in a parallel system. Scheduling parallel jobs is performed as a means to balance the load of a system in order to improve the performance of a parallel application. Parallel job scheduling is presented as a mapping between two graphs: one represents the dependency of jobs and the other represents the interconnection among processors. The usual implementation of parallel job scheduling algorithms is via the master-slave paradigm. The master-slave paradigm has inherent communication bottleneck that reduces the performance of the system when more processors are needed to process the jobs. The multiagent approach attempts to distribute the communication latency among the processors which improves the performance of the system as the number of participating processors increases. Presented in this paper is a framework for the behavior of an autonomous agent that cooperates with other agents to achieve a community goal of minimizing the processing time. Achieving this goal means an agent must truthfully share information with other agents via {\em normalization}, {\em task sharing}, and {\em result sharing} procedures. The agents consider a parallel scientific application as a finite-horizon game where truthful information sharing results into performance improvement for the parallel application. The performance of the multiagent-based algorithm is compared to that of an existing one via a simulation of the wavepacket dynamics using the quantum trajectory method (QTM) as a test application. The average parallel cost of running the QTM using the multiagent-based system is lower at higher number of processors.
\end{abstract}

\section{Introduction}\label{sec:1}

Parallel job scheduling is a problem of assigning a set~$J=\{j_i|i=0,1,\dots,n-1\}$ of parallel jobs onto a set~$P=\{p_j|j=0,1,\dots,m-1\}$ of parallel processors such that certain criteria are met. A major criterion is to minimize the overall parallel execution time~$T_P$ for~$J$. An ideal solution to minimizing~$T_P$ is to assign $n$~processors to $n$~jobs (i.e., $m=n$). When using $n$~processors, the overall execution time for~$J$ is $T_P=\max(t_i)$, where $t_i$ is the processing time for $j_i, \forall i$. However, for parallel scientific applications, the usual $n=|J|$ ranges from tens of thousands to several hundreds of millions. Setting $m=n$ is not a practical solution and thus, $m << n$. 

The simplest scheduling scenario is when the jobs' processing times are the same (i.e., $t_0=t_1=\cdots=t_{n-1}$) and the $m$ processors have the same processing speed (i.e., processors are homogeneous). The trivial solution to this scenario is to assign $\lceil n/m\rceil$ jobs to each processor. Here, the overall execution time simply becomes $T_P=\lceil n/m\rceil\times t_0$. However, in real-world problems, the job processing times are not the same and the processors' computing powers are different (i.e., processors are heterogeneous). The real-world situation reduces the scheduling problem into a {\em modified bin-packing problem} where the jobs are mapped as the objects to be packed while the processors are the bins. In the modified bin-packing problem, given are $n$~objects of different sizes to be distributed into $m$~bins of different capacities such that there is an {\em equitable} object distribution. In the modified 3D bin-packing problem for example, the objects are of different weights while the bins are of different volume. Equitable in this scenario means that the total weight of the objects in one bin is approximately equal to that of the other bins. The optimal solution to the scheduling problem can be obtained using a solution to the bin-packing problem. In scientific parallel applications, however, the processing times of jobs are not known prior to processing. Therefore, heuristics have been developed to dynamically assign $n$~jobs to $m$~processors such that the criterion of minimizing~$T_P$ is met.

In recent years, research advances in dynamic scheduling of parallel jobs have contributed to improving the performance of scientific applications under heterogeneous processors. Dynamic job scheduling schemes such as \emph{Factoring}, \emph{Fractiling}, \emph{Weighted Factoring}, \emph{Adaptive Weighted Factoring}, and \emph{Adaptive Factoring} have been proposed and successfully implemented in the  parallelization of Monte-Carlo simulations~\cite{hummel92}, many-body simulations~\cite{banicescu95,banicescu98}, radar applications~\cite{hummel96}, computational fluid dynamics~\cite{banicescu02,banicescu03}, and simulation of wavepacket dynamics using the quantum trajectory method~\cite{carino03}. These scheduling schemes are based on probabilistic analyses that take into consideration other variabilities that can make the problem complicated when solutions are sought in real time. The variabilities are brought about by systemic factors such as the variance in processor performance during execution and the variance in network latency. The performance of the processors might vary during job processing, even when the processing speed is known. For example, the performance of a processor might be affected when a server daemon is woken up or when the processor is interrupted by the hardware. 

The above mentioned job scheduling schemes greatly improve the performance of scientific parallel applications. However, these schemes do not come without inherent problems. For parallel applications utilizing $m$~processors, these schemes are implemented using a {\em master-slave} strategy where one processor acts as a master and scheduler. The master processor dynamically assigns jobs to slave\footnote{The master is also a slave.} processors following a scheduling policy defined by the scheme. For example, in Fractiling, the scheduling policy is based on the characteristics of fractals. After being assigned with a {\em chunk} of jobs, the performance of the slaves  is evaluated by the master. A chunk is some number of jobs~$c$ such that $c<<\lceil n/m\rceil$. The future assignment of new chunks for the slave processor is computed based on its performance of the previous chunk assignment. The performance measure is simply the time it took for the slave to process a given chunk. When a slave is done with its assigned chunk, the slave communicates with the master its chunk processing time. The master then communicates to the slave the next chunk of jobs that the slave needs to process. The size of the next chunk is defined by the scheduling policy. If all $m$~jobs have already been assigned and one slave processor finishes early than the other, the master asks the slowest processor to give up {\em some} of its jobs to the now idle slave processor. The number of jobs that are to be given up is again based on the scheduling policy defined by the scheme. 

The inherent problem with the master-slave strategy is that when more processors are required by the parallel application, the number of slaves that a master has to communicate with increases. The benefits of the dynamic scheduling schemes are greatly offset by the latency due to communication bottleneck with the master, compounded with the type of network interconnect the processors are using. When more slaves communicate to the master, the master is overwhelmed with communication requests and might not be able to process jobs assigned to itself. The assigned chunk of jobs to the master will eventually be reassigned to other slaves requiring more communication. This scenario leaves one processor doing only scheduling rather than doing job processing. The communication latency greatly reduces the performance of the application when~$m$ is increased.

The problem with master-slave strategy is that only one processor decides for the rest. All the slave processors are dependent on the master processor. The slave processors are not responsible for scheduling and are only tasked to process the assigned jobs and communicate to the master their performance and status. The master on the other hand is tasked to process its assigned chunk of jobs while it keeps record of the slaves' status. The master also runs the scheduling policy and communicates to $m-1$~slaves\footnote{The master does not incur communication latency when it communicates to itself.}. The master-slave strategy can be seen as a multiagent system where the master is an independent agent while the slaves are dependent yet specialized agents\footnote{The slave's specialization is to process jobs only.}.  If all processors will instead act as independent yet truthful agents that follow the same scheduling policy, the communication latency brought about by communicating to only one master will be evenly distributed to the other $m-1$~processors. Thus, a multiagent approach to scheduling parallel task might be a better strategy than the master-slave one.

In this effort, the scheduling scheme for parallel tasks is mapped into a multiagent strategy. Here, processors are considered autonomous agents with a self-interested goal of finishing its assigned chunk of jobs at the earliest time. However, the agents are members of a community with a social goal of finishing~$J$ at the shortest possible time~$T_P$. The agents are independent in the sense that it can decide for itself. However, the agents are still dependent on other agents such that the social goal of minimizing~$T_P$ is obtained. This dependency requires other agents to communicate with other agents. However, since agents are now independent, they are required to truthfully communicate their status and performance such that the other agents will have a global perspective of the community. This global perspective will help the agents come up a schedule for themselves such that both their personal goals and the community goals are achieved.

This paper presents a multiagent framework for parallel task scheduling. Section~\ref{sec:3} presents a formal definition of the scheduling problem, reviews some parallel job scheduling schemes, and presents the general structure of parallel scientific applications. Section~\ref{sec:4} presents the solution to the scheduling problem using a multiagent strategy. The behavior of an agent will be defined while the parallel scientific application will be considered as a finite-horizon game for the agents. Section~\ref{sec:4b} discusses and presents the results of the experiments performed to assess the performance of a test application when running adaptive factoring and multiagent-based algorithms. Section~\ref{sec:5} gives a summary of the paper.

\section{The Parallel Job Scheduling Problem}\label{sec:3}

\subsection{Parallel Jobs}

Parallel jobs in a scientific application usually come as iterates of a parallel loop. In a parallel loop, the iterates are independent of each other and therefore can be parallelized. For example, the code fragment in Figure~\ref{fig:parloop} shows a parallel loop with iterates that are not dependent on the other iterates. In this fragment, the computation of $j[i]$ is not dependent on $j[i-k]$, $\forall k, 0\leq k < i-1$.

\begin{figure}[ht]
\centering
\begin{verbatim}



   /* some sequential code fragments here */
   for i = 0 to n-1 do
   begin
     j[i] = f(data[i]);
   end;
   /* more sequential code fragments here */
\end{verbatim}
\caption{A code fragment for a parallel loop.}
\label{fig:parloop}
\end{figure}

An example of a scientific application with parallel loops is the simulation of the three-dimensional many-body system. The many-body simulation is a real-world application common to the physical science discipline such as molecular dynamics, astrophysics, plasma physics, and many others. It simulates the evolution of particles in three dimensions given the initial positions and velocities of the particles and experiencing gravitational forces due to their interaction with one another. Their interactions are governed by the Inverse Square Law and the net force acting on a particle determines the particle's next position and velocity. The naive algorithm for many-body simulation has time complexity of $O(N^2)$, where $N$ is the number of bodies. An algorithm, termed Fast Multipole Algorithm (FMA) was developed that improved the time complexity to $O(N)$~\cite{greengard87}. The parallel implementation of FMA uses an oct-tree structure that represent a hierarchical decomposition of the three dimensional space. The FMA recursively divides the space into cubical boxes, which are ordered hierarchically in a tree. In FMA, the effect of a particle at a short distance is represented directly, while at longer distances, the effect is pooled with other particles and represented by a multipole expansion. The expansion is transfered to distant boxes and expressed as Taylor expansions. The directed acyclic graph (DAG) representation of the recursive division of space has a very high degree of concurrency and is therefore amenable to parallelization.

\subsection{Graph Representation of Parallel Jobs}

\begin{figure}[ht]
\centering
\epsfig{file=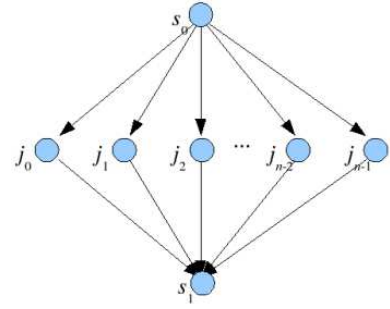, width=2.0in,height=1.62in}
\caption{A directed acyclic graph abstraction of a parallel loop.}
\label{fig:dag}
\end{figure}

The independent iterates in a parallel loop shown in Figure~\ref{fig:parloop} can be represented by a DAG $G_J(J \bigcup \{s_0, s_1\}, E_J)$ with $n+2$ vertices and an edge set $E_J$(Figure~\ref{fig:dag}). The $n$~vertices $j_i\in J, \forall i=0,1,\dots, n-1$, represent the $n$~iterates while the extra vertices $s_0$ and $s_1$ represent the sequential portions before and after the parallel loop of the code fragment in Figure~\ref{fig:parloop}. With the DAG, the maximum degree of parallelization can be visually seen. In Figure~\ref{fig:dag} for example, the maximum degree of parallelization is~$n$ for a DAG abstraction of a loop with $n$~independent iterates.

In most scientific applications, the parallel loops maybe embedded in a time-stepping loop. An example code fragment of a time-stepping loop with embedded parallel loops is shown in Figure~\ref{fig:time-step1}. In this code fragment, the iterates may now be dependent on the computations of the previous time-step. Thus, under these applications, the processors may only process jobs one time-step at a time.

\begin{figure}[ht]
\centering
\begin{verbatim}
   for t=0 to MAXTIME-1 do
   begin
     /* some sequential code fragments here */
     for i = 0 to n-1 do
     begin
       j[t][i] = f(data[t-1][i]);
     end;
   /* more sequential code fragments here */
   end;
\end{verbatim}
\caption{A code fragment for a parallel loop in a time-stepping scientific application.}
\label{fig:time-step1}
\end{figure}

\subsection{Graph Representation of Parallel\\Processors}

\begin{figure}[ht]
\centering
\epsfig{file=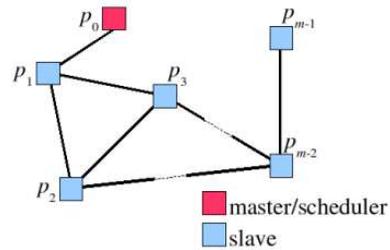, width=2.0in,height=1.36in}
\caption{An undirected graph abstraction of a network of parallel processors.}
\label{fig:network}
\end{figure}

To process the parallel jobs~$J$ in a cost-optimal manner, $m<<n$ processors will be needed such that the parallel processing time~$T_P$ for~$J$ is minimized. The $m$~processors are connected in a network and their interconnection is represented by an undirected graph $G_P(P, E_P)$ with $m$~vertices and an edge set $E_P$. The edges in $E_P$ represent the direct interconnection between two processors. Figure~\ref{fig:network} shows an undirected graph representation of a general interconnection network for a parallel system.

\subsection{The Scheduling Problem}

\begin{figure}[ht]
\centering
\epsfig{file=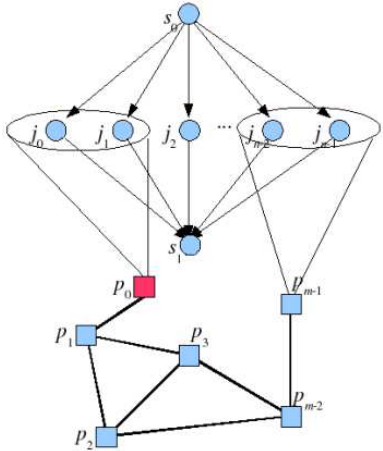, width=2.0in,height=2.39in}
\caption{Mapping of two graphs as an abstraction for scheduling.}
\label{fig:mapping}
\end{figure}

With the DAG representation of the parallel jobs $G_J$ and the graph representation of the interconnect network for the processors $G_P$, the scheduling problem reduces into a mapping between $G_J$ and $G_P$ subject to some constraints. The scheduling problem finds an assignment of the vertices\footnote{The extra 2 vertices in the DAG that represent the sequential code fragments are not included in the assignment.} in~$G_J$ to the vertices in~$G_P$. Usually, the constraint is to maximize data locality. Maximizing data locality means assigning jobs to processors such that communications with far away processors are avoided. Figure~\ref{fig:mapping} presents the abstraction of a parallel job scheduling as a graph mapping.

The simplest mapping is to obtain the simplest scenario described in Section~\ref{sec:1} above (i.e., $m=n$). However, for practical reasons, $n$~processors cannot be utilized. In most scientific applications, the number of iterates in a parallel loop ranges from tens of thousands to tens of millions, thus using $n$~processors is not cost-optimal. Moreover, the processing time for the parallel jobs cannot be obtained {\em a priori}. The bin-packing solution for this problem cannot be utilized. The complexity of the scheduling problem is compounded with the use of heterogeneous processors and the unpredictability of the system and of the network. For example,  interruptions to processing such as server daemons waking up or hardware interrupts cannot be predicted beforehand even if the processors computing speeds are known {\em a priori}, or via some profiling during processing time. It is also possible that the graph representation of the interconnection of processors will change during processing. In a grid environment for example, processors might suddenly go offline. Thus, scheduling of parallel jobs to processors can only be done online via some dynamic scheduling algorithms.

\subsection{Dynamic Scheduling Algorithms}\label{sec:DLB}

In systems that use heterogeneous processors, the difference in processor speeds and memory capacities coupled with the dynamism in processor interconnection can significantly impact the performance of scientific applications. For example, load and resource availability of processors in a grid is unpredictable, and thus it is difficult to know in advance what the effective speed of each machine would be. For effectively load balancing scientific applications, algorithms that derive from theoretical advances in research on scheduling parallel jobs with variable processing times have extensively been studied \cite{kruskal85,polychronopoulos87,tzen93,hummel92,markatos94}. As a result, dynamic job scheduling algorithms based on a probabilistic analysis have been proposed, and successfully implemented in a cluster environment for a number of scientific applications \cite{hummel92,banicescu95,hummel96,banicescu98,banicescu02,banicescu03,carino03,carino04}.

An example of a dynamic scheduling scheme is the one that uses a scheduling policy based on {\em Factoring}~\cite{hummel92}. In this scheme, a probabilistic analysis allows the formulation of {\em factoring rules}. Jobs are assigned to processors in chunks of variable sizes following the factoring rules. The selection of chunk sizes requires that they have a high probability of being completed by the processors before the optimal time. The chunk sizes are dynamically computed during the execution of the application. The larger chunks have relatively little overhead, and their unevenness can be smoothed over by the smaller chunks. Another scheduling policy that improves Factoring is {\em Fractiling}. It combines the scheduling technique that balances processor loads with data locality by exploiting the self-similarity properties of fractals. This method has successfully been implemented for many-body simulations on distributed shared address space and message-passing environments~\cite{banicescu95,banicescu98}.

Other dynamic scheduling schemes that have been successfully implemented and found to improve the performance of Factoring are {\em Weighted Factoring}, {\em Adaptive Weighted Factoring}, and {\em Adaptive Factoring}. The adaptive factoring scheme is a more general model for scheduling. In specific conditions of processor speed and job workloads, the Adaptive Factoring converts into the other schemes. These schemes are described in details elsewhere~\cite{hummel96,banicescu03,banicescu02} and will not be discussed in this paper. Suffice it to say that all these schemes are implemented in a cluster environment using a master-slave strategy which,  as discussed in Section~\ref{sec:1}, has inherent communication bottleneck for higher~$m$.

\section{Multiagent Approach to\\Scheduling}\label{sec:4}

The master-slave strategy can be regarded as a special case of a multiagent system. The master is an agent that has total {\em control} of the $m-1$~slaves. Control here means that the master has all the decision making responsibilities for the community composed of itself and the slaves. Since the master itself is the scheduler, it needs to have an updated global information of the status of the system. This means that all slaves will have to periodically\footnote{The period here depends on the scheduling policy being used.} communicate to the master. The over head due to communication increases with the number of slaves present in the community. Thus, at higher values of~$m$, the scheduling schemes with master-slave strategy tend to reduce its performance.

In this section, a paradigm shift for designing and analyzing scheduling algorithms  will be discussed. This shift is from purely master-slave paradigm to multiagent paradigm. Here, multiagent means a system composed of autonomous agents, which is in contrast from the master-slave paradigm where only the master is autonomous. The main idea here is that if all agents are autonomous, then the communication bottleneck inherent to the master-slave strategy can somewhat be distributed among the agents. Therefore at higher~$m$, performance of the parallel application will vastly improve.

\subsection{Processors as Agents}

In the multiagent approach to scheduling, the processors are considered as autonomous agents that belong to a community. The community is tasked to process a set~$J$ of $n$~parallel jobs. Initially, the agents do not know the respective computing speed of the other agents. However, the number of agents that belong to the community and the respective address of each agent are common knowledge. The agents may use the same addressing scheme as the processors' (i.e., $P=\{p_0, p_1, \dots, p_{m-1}\}$). This addressing scheme is arbitrary and does not favor any agent. What is important here is that every agent knows its address as well as the other agents'. 

\subsection{Initial Job Assignment}

The initial job assignment is guided by an assumption that the processing times of all the jobs in~$J$ are the same. Similarly, processors are assumed to be homogeneous. These assumptions are not common knowledge for the agents but just a simplistic way of initializing the system with a job partition to start with. The $n$~jobs are equally divided to $m$~agents and each agent receives $\lceil n/m \rceil$ jobs. A trivial assignment would be such that agent~$p_0$ is assigned to jobs $j_0$ to $j_{\lceil n/m \rceil-1}$, agent~$p_1$ to jobs $j_{\lceil n/m \rceil}$ to $j_{2\lceil n/m \rceil-1}$, and so on. If the jobs will come from a single processor, then a {\em scatter} or {\em one-to-all personalized communication}~\cite{grama03} may be utilized. After initialization, the agents will perform a {\em task sharing} or {\em task passing} strategy~\cite{durfee01} based on a prediction to job processing times.

\subsection{Task and Result Sharing with\\Homogeneous Agents}

Each agent will process the jobs assigned to it using some statistical sampling order and size. The goal of sampling is for the agent to predict the profile of the processing time of the $\lceil n/m \rceil$ jobs assigned to it using some sample size $s << \lceil n/m \rceil$. For example, agent $p_0$ will process jobs $\{j_i | i=0, \lceil n/m \rceil / s, 2\lceil n/m \rceil / s, \dots \}$. Based on $p_0$'s processing times on the sampled jobs, the processing times for all the remaining jobs assigned to it can be predicted via a curve fitting or interpolation technique (e.g., nonlinear regression, divided difference, etc.). After some time, all agents will have a local knowledge of the predicted processing times of the remaining jobs assigned to them. If all agents will truthfully exchange this information to all other agents (i.e., via an {\em all-to-all broadcast}~\cite{grama03}), then each of them will have a global knowledge of the predicted processing times of the remaining jobs assigned to the community. With this information, the modified bin-packing algorithm may be employed by each agent to reallocate jobs to processors taking in consideration the jobs that already have been sampled. This information sharing is what is termed as {\em result sharing} by Durfee~\cite{durfee01}.

\subsection{Normalization Procedure with\\Heterogeneous Agents}

In the case of heterogeneous processors, each agent will need to perform a normalization procedure to account for the variability in processor speeds. The normalization involves sampling a job assigned to other agents and which already have been sampled by that agent. For example, if agent $p_0$ found job $j_0$ to have a processing time of $t_{0,0}$, and if agent $p_1$ processed the same job $j_0$ and found it to have a processing time of $t_{0,1}$, then the ratio between $t_{0,0}$ and $t_{0,1}$ provides the relative speed of $p_1$ to $p_0$, and {\em vice versa}. If all agents will perform this normalization procedure and will truthfully communicate the information from this procedure to other agents, then all agents will have the global information of their relative speeds compared to the others. An all-to-all broadcast may again be utilized to efficiently exchange the information. Based on this information, the modified bin-packing algorithm for uneven bins may be utilized by each agent to come up with a global reallocation of jobs. Since all agents will have the same solution, when agent $p_i$ asks for a job from agent $p_j$, $p_j$ already knows what job to give and must freely give it. The agent behavior will be guided by some payoff scheme that is a function of the performance of the community when solving the assigned jobs. The detailed discussion of the payoff scheme will be presented in the future.

\subsection{Time-stepping Application as a Game}

\begin{figure}[hb]
\centering
\epsfig{file=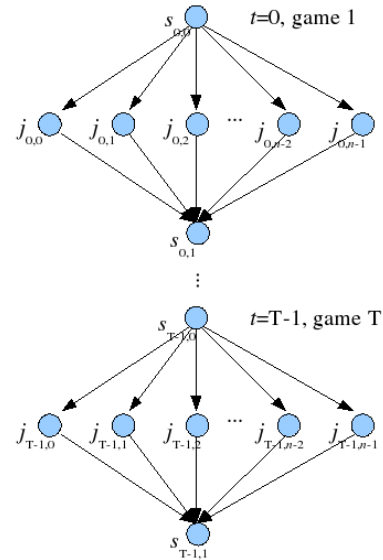, width=2.0in,height=2.94in}
\caption{A DAG representation of a time-stepping application regarded by agents as a finite-horizon game.}
\label{fig:time-step}
\end{figure}

A code fragment of a scientific application presented in Figure~\ref{fig:time-step1} presents parallel loops embedded in a time-stepping loop. If the time-stepping loop iterates from 0 to some $t_{\rm MAXTIME}-1$, then the application maybe regarded by the $m$~agents as a finite-horizon game with $t_{\rm MAXTIME}$ horizons (Figure~\ref{fig:time-step}). 
Each game, the agents will cooperatively process $n$~jobs via normalization, task sharing and result sharing such that the community goal is achieved. The game is formalized by a mechanism design such that the Nash equilibrium is a strategy where agents must truthfully share information and cooperate with other agents. The formalization of mechanism design for the {\em game theoretic} perspective of a time-stepping scientific application will be discussed and presented in the future.

\begin{figure*}[hbt]
\centering
\begin{verbatim}
Initialize positions r[], velocities v[], and probability densities p[]
for time_step=0 to T-1 do
begin
    for pseudoparticle i = 1 to n (Loop 1)
    begin
       MWLS(i, r[], p[], np1, nb); compute quantum potential Q[i];
    end
    for pseudoparticle i = 1 to n (Loop 2)
    begin
       MWLS(i, r[], p[], np2, nb); compute quantum force f_q[i];
    end
    for pseudoparticle i = 1 to n (Loop 3)
    begin
       MWLS(i, r[], p[], np2, nb); compute derivative of velocity dv[i];
    end
    for pseudoparticle i = 1 to n (Loop 4)
    begin
       compute classical potential V[i], classical force f_c[i];
    end
    Output t, r[], v[], p[], V[], f_c[], Q[], f_q[], dv[];
    for pseudoparticle i = 1 to n (Loop 5)
    begin
       update p[i], r[i], v[i];
    end
end
\end{verbatim}
\caption{Simulation of wave packet dynamics using the QTM.}
\label{fig:qtm}
\end{figure*}

\section{Performance Evaluation and Results}\label{sec:4b}

In this section, the test application and evaluation methods employed to assess the performance of the multiagent-based over adaptive factoring are described. The results of the evaluation are then presented, analyzed, interpreted and discussed.

\subsection{The Quantum Trajectory Method}\label{sec:QTM}

The test application used in this study is the simulation of wave packet dynamics using the quantum trajectory method (QTM). QTM is a numerical solution to the time-dependent Schr\"{o}dinger equation (TDSE)
\begin{equation}
  \imath h\frac{\partial}{\partial t}\Psi = \mathcal{H}\Psi, \quad\Psi = - \frac{h^2}{2\mu}\nabla^2+V
\end{equation}
TDSE describes the dynamics of quantum-mechanical systems composed of a particle of mass~$\mu$ moving in a potential~$V$. A set of $n$ pseudoparticles, each of mass~$\mu$, is deployed to represent the physical particle. Each pseudoparticle executes a {\em quantum trajectory} governed by the Lagrangian equations of motion and the quantum potential~$Q$. Derivatives of the probability density~$\rho$, $Q$, and the velocity~$v$ used for updating the equations of motion are obtained by curve-fitting the numerical values of these variables using a moving weighted least squares (MWLS) algorithm, and analytically differentiating the least squares curves~\cite{lopreore99}.

The QTM proceeds as outlined in Figure~\ref{fig:qtm}. The MWLS parameters~$np$ and~$nb$ represent the dimensions of the least-squares matrix to be solved. The loops over pseudoparticles are parallel loops. Loops~1 through~3 are computationally bound, consuming the bulk of the CPU time. Due to adaptivity in the MWLS, these loops have nonuniform iterate execution times, thus, dynamic scheduling of the iterates is necessary for load balancing~\cite{carino04}.

\subsection{Experimental Setup and Results}

The QTM was used to assess the performance of the multiagent-based scheduling over the adaptive factoring scheduling. Due to limited space, the comparison with the other scheduling algorithms will be presented in the future. All runs were conducted on an adhoc general-purpose cluster at the Institute of Computer Science, University of the Philippines Los Ba\~{n}os. The adhoc cluster is a 64-node, 64-processor cluster with a combined memory of 32 GB. The processors are of Pentium III, Pentium IV and AMD Athlon architectures with processing speeds that range from 800MHz to 2.4GHz. They are interconnected through 10 Mbps Ethernet switches that are interconnected through a 100 Mbps Ethernet uplink. Since this cluster is a general-purpose, adhoc cluster, the network traffic volume is not predictable (i.e., other jobs maybe running along with the QTM simulations). 

\begin{figure*}[ht]
\centering
\epsfig{file=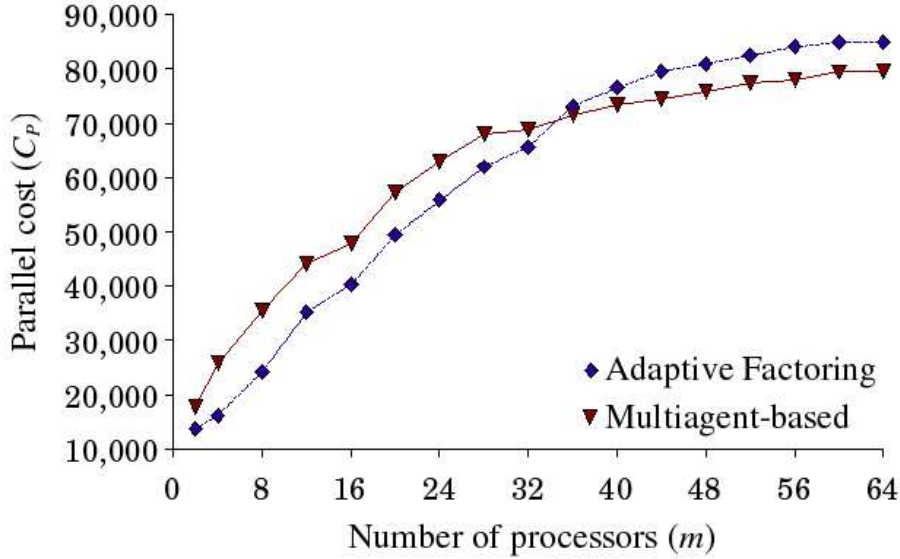, width=4.75in,height=3in}
\caption{Average parallel cost for wave packet simulation with 501 pseudoparticles using adaptive factoring and multiagent-based scheduling at increasing number of processors.}
\label{fig:result}
\end{figure*}

A wave packet of 501 pseudoparticles was simulated for 10,000 time steps. The numerical characteristics of the QTM preclude wave packet sizes beyond a few thousand pseudoparticles; hence, the wave packet size cannot be arbitrarily increased to generate larger problems for execution on more processors. The performance metrics measured was the parallel cost $C_P=m \times T_p$. Each simulation run was replicated five times to model the variability of $C_P$ induced by the underlying computing architecture. 

The average parallel costs for running the QTM using the adaptive factoring and the multiagent-based scheduling are graphed in Figure~\ref{fig:result}. The graph indicates that:
\begin{enumerate}
\item The adaptive factoring is, on the average, cost-effective at lower $m$. In fact, the adaptive factoring is less costly at $m\leq 32$.
\item The multiagent-based scheduling has lower average parallel cost at higher $m$. Graph shows that at $m>32$, the average parallel cost for multiagent-based scheduling is lower than that of the adaptive factoring.
\end{enumerate}

\section{Summary}\label{sec:5}

This paper presents a framework for a dynamic solution to parallel job scheduling using the multiagent paradigm. The purpose of job scheduling is to balance the load to improve the performance of a parallel application. The scheduling problem is presented as a mapping between the DAG representation $G_J$ of the set of jobs~$J$ and the graph representation $G_P$ of the set of processors~$P$ in an arbitrary interconnect network. The static and offline solution to the scheduling problem can be optimally obtained using the bin-packing algorithm for tractable values of~$|J|$ and~$|P|$. Due to variability in the system, and because the processing times for the jobs in~$J$ are not known {\em a priori}, the scheduling problem can only be solved online. Most known dynamic scheduling schemes use a master-slave strategy that is inherently susceptible to communication bottleneck when~$|P|$ is increased. The multiagent approach regards processors as agents. The approach reduces the bottleneck by distributing the communication to other agents. All agents in the community must truthfully share information via normalization, task sharing, and result sharing. The information exchange may be efficiently performed using the one-to-all personalized communication and the all-to-all broadcast. The behavior of the agents is defined by a payoff function which will be discussed in the future. A time-stepping application can be regarded by agents as a finite-horizon game. The game formalization is a mechanism design that has Nash Equilibriums for cooperation and truthful information sharing. The mechanism design formulation will be presented in the future. The performance of the algorithm is compared to the performance of the adaptive factoring scheme using the QTM as a test application. When QTM is simulating 501 pseudoparticles, the multiagent-based scheduling is less costly, on the average, at $m>32$ compared to the adaptive factoring. Thus, at higher~$m$, the multiagent-based scheduling outperformed the adaptive factoring scheme.

\section{Acknowledgments}

The author thanks Dr. Eric Hansen of the Department of Computer Science and Engineering, Mississippi State University (MSU), Mississippi State, MS 39762, USA for his valuable theoretical inputs about multiagent systems. The author also appreciates the assistance of Dr. Ricolindo L. Cari\~{n}o of the Center for Computational Sciences, Engineering Research Center, MSU in providing him the parallel codes for the dynamic scheduling algorithms discussed in Section~\ref{sec:DLB} and for the QTM discussed in Section~\ref{sec:QTM}. The author would also like to thank the Institute of Computer Science, University of the Philippines Los Ba\~{n}os for its support of this work through the use its computer and research laboratories.

\bibliographystyle{plain}
\bibliography{pcsc06}
\balancecolumns
\end{document}